\documentclass[conference]{IEEEtran}
\usepackage{ifpdf}

\usepackage{cite}

\ifCLASSINFOpdf
  \usepackage[pdftex]{graphicx}
  \graphicspath{../figs/}
  \DeclareGraphicsExtensions{.pdf,.jpeg,.png}
\else
  \usepackage[dvips]{graphicx}
\fi
\usepackage{amsmath, amssymb}
\usepackage{algorithm, algorithmic}

\usepackage{array}
\usepackage{fixltx2e}

\usepackage{stfloats}
\usepackage{url}

\usepackage[colorlinks,
linkcolor=red,
anchorcolor=green,
citecolor=blue]{hyperref}

\usepackage{multirow}
\usepackage{booktabs}
\usepackage{arydshln}
\usepackage[table,xcdraw]{xcolor}
\usepackage{longtable}
\usepackage{booktabs}
\usepackage{threeparttable}
\usepackage{lscape}
\usepackage{subfigure}
\usepackage{threeparttable}
\usepackage{verbatim}
\usepackage{marvosym}
\usepackage{algorithmic}
\usepackage{algorithm}
\usepackage{pifont}

\newif\ifarxiv
\arxivtrue
\usepackage{lastpage}
\usepackage{fancyhdr}
\fancypagestyle{firstpage}
{
	\fancyhead{}
	
	\ifarxiv
	\fancyfoot[C]{\scriptsize \copyright~2024 IEEE. Personal use of this material is permitted. Permission from IEEE must be obtained for all other uses, in any current or future media, including reprinting/republishing this material for advertising or promotional purposes, creating new collective works, for resale or redistribution to servers or lists, or reuse of any copyrighted component of this work in other works. 
		This work has been accepted at the 29th Asia and South Pacific Design Automation Conference (ASP-DAC 2024).}
	\else
	\fi
}

\newcommand{\rTwoFC}[1]{\textcolor{black}{#1}}
\newcommand{\rOneFC}[1]{\textcolor{black}{#1}}
\newcommand{\rOneLQ}[1]{\textcolor{black}{#1}}
\newcommand{\rtwoLQ}[1]{\textcolor{black}{#1}}
\newcommand{\rOneHLW}[1]{\textcolor{black}{#1}}
\newcommand{\rTwoHLW}[1]{\textcolor{black}{#1}}
\newcommand{\rOneLIN}[1]{\textcolor{black}{#1}}

\newcommand{\Review}[1]{\textcolor{black}{#1}}

\begin{document}


%

\title{A Precision-Scalable RISC-V DNN Processor with On-Device Learning Capability at the Extreme Edge}

\author{
    Longwei~Huang, Chao~Fang, Qiong~Li, Jun~Lin$^{(\textrm{\Letter})}$, Zhongfeng~Wang$^{(\textrm{\Letter})}$ \\
	\IEEEauthorblockA{
		School of Electronic Science and Engineering, 
		Nanjing University, China\\
		Email:
        \{522022230032, fantasysee, qiongli\}@smail.nju.edu.cn, \{jlin, zfwang\}@nju.edu.cn
    }
}

\maketitle
\thispagestyle{firstpage}


\begin{abstract}
\rOneHLW{Extreme edge platforms, such as in-vehicle smart devices, require efficient deployment of quantized deep neural networks (DNNs) to enable intelligent applications with limited amounts of energy, memory, and computing resources.}
However, many edge devices struggle to boost inference throughput of various quantized DNNs due to the varying quantization levels, and these devices lack floating-point (FP) support for on-device learning, which prevents them from improving model accuracy while ensuring data privacy.
\rOneLIN{To tackle the challenges above, we propose a precision-scalable RISC-V DNN processor with on-device learning capability.}
\rOneLIN{It facilitates diverse precision levels of fixed-point DNN inference, spanning from 2-bit to 16-bit, and enhances on-device learning through improved support with FP16 operations.}
Moreover, we employ multiple methods such as FP16 multiplier reuse and multi-precision integer multiplier reuse, along with balanced mapping of FPGA resources, to significantly improve hardware resource utilization.
Experimental results on the Xilinx ZCU102 FPGA show that our processor significantly improves inference throughput by 1.6$\sim$14.6$\times$ and energy efficiency by 1.1$\sim$14.6$\times$ across various DNNs, compared to the prior art, XpulpNN. Additionally, our processor achieves a 16.5$\times$ higher FP throughput for on-device learning.
\end{abstract}

\section{Introduction} \label{sec:intro}
\rOneFC{
Extreme edge platforms as Internet-of-things (IoT) nodes, such as in-vehicle and wearable smart devices, are confronted with significant challenges of constrained power supply, memory space, and computing resources, but there is a growing need to deploy computation-intensive tasks leveraging deep neural networks (DNNs) on these platforms.
Hence, deploying quantized DNN models \cite{wu2016quantized, liu2021post, ding2018quantized, tian2023bebert, zhou2018explicit} emerge as an efficient solution at the extreme edge due to their substantial reductions on energy, storage, and computation costs.
}

\rOneFC{Nevertheless, the efficient deployment of quantized DNNs on extreme edge platforms presents several challenges.}
\rOneFC{Firstly,} quantized DNNs \cite{wu2016quantized, liu2021post, ding2018quantized, tian2023bebert, zhou2018explicit} come in various precisions, ranging from 16-bit integers (INT16) to 2-bit integers (INT2), catering to different targeted applications. However, many processors \cite{yang2021flexacc, suda2016throughput, qiu2016going, chang2021mix, guo2017angel, tortorella2022redmule} are unable to effectively perform precision-scalable operations, hindering the potential throughput improvement for these quantized DNNs.
\rOneHLW{Secondly, extreme edge platforms require on-device learning capability\cite{cai2020tinytl} with floating-point (FP) precision to enhance model accuracy and preserve data privacy. However, existing precision-scalable processors \cite{sun2022film, ottavi2023dustin, askarihemmat2023barvinn, he2023agile} lack support for FP operations, making them incapable of realizing on-device learning.}
Finally, the resource requirements of extreme edge applications vary significantly, underscoring the importance of employing configurable hardware architectures.
\rOneFC{In this case, FPGA devices \cite{wu2021flexible, fang2021accelerating, wang2019low} emerge as a promising alternative to CPUs and GPUs because of their superior energy efficiency, cost-effectiveness, and ability to be customized for specific applications.}
\rOneFC{However, deploying BARVINN \cite{askarihemmat2023barvinn} and XpulpNN \cite{garofalo2021xpulpnn} on FPGA devices yields sub-optimal utilization of hardware resources, restricting the potential for achieving higher throughput in extremely edge quantized DNN applications.}
In summary, existing FPGA-based processors at the extreme edge requires three key improvements: supporting precision-scalable inference for various precision quantized DNNs, enabling \Review{floating-point} computation for on-device learning, and increasing resource utilization on FPGA for higher throughput.

Therefore, in this paper, we propose a precision-scalable RISC-V DNN processor with on-device learning capability at the extreme edge \rOneFC{using FPGA devices}.
\rTwoFC{Specifically, we design a re-configurable DNN processor that integrates a tightly-coupled co-processor into a PULP~\cite{conti2016pulp} cluster.}
\rOneFC{The co-processor achieves a significant throughput improvement for DNNs by employing systolic computing dataflow on resource-efficient precision-scalable processing elements (PEs).}
\Review{The proposed PE supports various types of fixed-point inference for diverse quantized DNNs with precision ranging from INT16 to INT2, and enables FP16 operations that are widely used in on-device learning\cite{micikevicius2017mixed}.}
\rOneFC{In addition, we leverage multiple techniques to improve hardware resource utilization including FP16 multiplier reuse, multi-precision multiplier reuse, and balanced mapping of FPGA resources.}
\rTwoHLW{The main contributions of the paper are as follows: 
\begin{itemize}
\item [1)] Architecture of a high-throughput, energy-efficient and resource-efficient RISC-V DNN processor.
\rTwoFC{By implementing a co-processor that leverages resource-efficient precision-scalable PEs with efficient systolic computing dataflow, our processor significantly improves both inference throughput and energy efficiency by up to 14.6$\times$ at various DNN precisions compared to XpulpNN\cite{garofalo2021xpulpnn}.}
\item [2)] Design of resource-efficient precision-scalable PEs. 
\rTwoFC{By integrating several hardware reuse techniques, we achieve a reduction of 25.8\% and 7.9\% in LUT and DSP overhead, respectively, for 8-bit precision-scalable multiplier trees. The saved resources can be used to add more PEs for further throughput improvement.}
\item [3)] Method of boosted on-device learning throughput along with increased resource utilization. 
\rTwoFC{By reusing INT16 multiplier of PEs to support FP16 operations, a maximum 16.5$\times$ on-device learning throughput improvement is achieved, while LUT and DSP utilization are increased by 4.7\% and 64.3\%, respectively, compared to XpulpNN.}
\end{itemize}}

\begin{figure*}[htbp]
    \centering
    \includegraphics[width=0.86\textwidth]{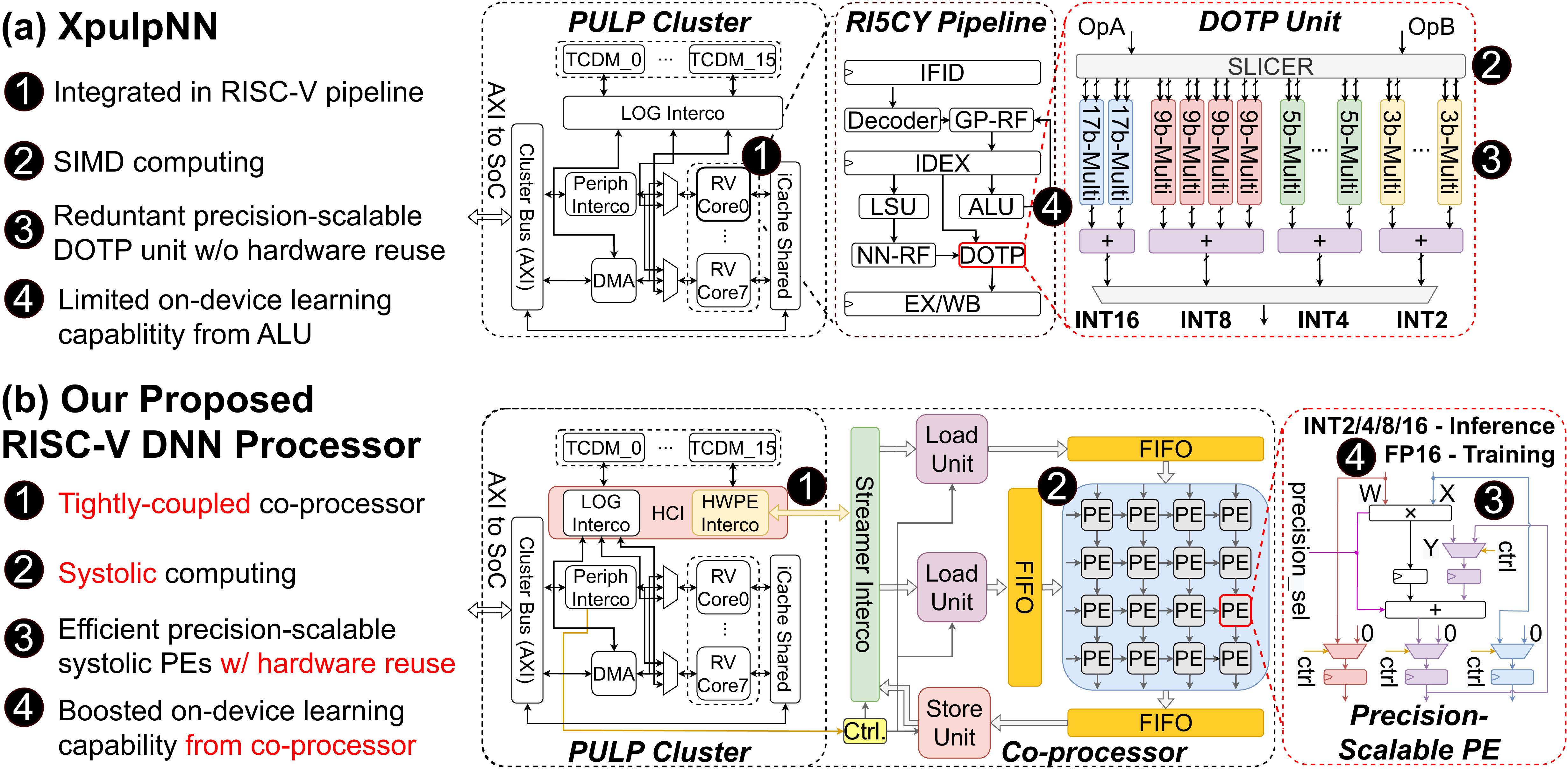}
    \caption{Comparison between the architecture of (a) XpulpNN \cite{garofalo2021xpulpnn} and (b) our proposed DNN processor.}
    \label{fig:system}
    \vspace{-0.3em}
    \vspace{-12pt}
\end{figure*}

\section{Related Works}\label{sec:bkg}

\rOneFC{Extreme edge processors are dominated by ARM or RISC-V instruction-driven microcontrollers. These devices face severe constraints on energy, memory, and computing, while confronting a substantial demand for deploying DNNs to realize intelligent applications.}
\rOneFC{To tackle these challenges, prior works \cite{wu2016quantized, liu2021post, ding2018quantized, tian2023bebert, zhou2018explicit} propose several dedicated DNN accelerators, but they sacrifice the general support for instruction-driven MCUs in extreme edge platforms across various IoT nodes. Moreover, \cite{chang2021mix, guo2017angel, suda2016throughput, qiu2016going} only support a single data type, limiting their effectiveness in deploying various quantized DNNs with precision ranging from INT2 to INT16 \cite{wu2016quantized, tian2023bebert, liu2021post, zhou2018explicit}.}

\rOneFC{In contrast, enhancing throughput of RISC-V processors for quantized DNNs \cite{askarihemmat2023barvinn, he2023agile, garofalo2021xpulpnn, ottavi2023dustin, yang2021flexacc, tortorella2022redmule} shows greater potential compared to exploiting dedicated DNN accelerators.}
\rOneFC{It maintains general support of RISC-V instructions, decouples software development and hardware implementation, and achieves significant performance in deploying DNNs.}
\rOneFC{Among them, FlexACC \cite{yang2021flexacc} and RedMulE \cite{tortorella2022redmule} enable only 8-bit or higher precision for DNN inference, which can hardly leverage the computational reduction potential of lower-bit quantized DNNs.} 
\rOneFC{The other RISC-V processors \cite{askarihemmat2023barvinn, he2023agile, garofalo2021xpulpnn, ottavi2023dustin} achieves significant speedup by supporting lower-bit DNN inference.}
\rOneFC{In addition, on-device learning at the extreme edge has gained attention for improving the accuracy of quantized DNNs while safeguarding data privacy.}
\rOneFC{Nevertheless, both BARVINN \cite{askarihemmat2023barvinn} and \cite{he2023agile} remove FP support, rendering them incapable of on-device learning. XpulpNN \cite{garofalo2021xpulpnn} and Dustin \cite{ottavi2023dustin} have weak on-device learning capability from the arithmetic logic unit (ALU) supporting FP operations in the RISC-V pipelines.}
\rOneFC{Compared with the above prior arts, our proposed RISC-V DNN processor not only efficiently supports precision-scalable quantized DNN inference varying from INT2 to INT16 but also significantly enhances on-device learning capability.}

\section{The Proposed RISC-V DNN Processor}\label{sec:design}

\subsection{Features of Our DNN Processor}\label{subsec:processor}
\rOneFC{PULP\cite{conti2016pulp} is an open-source RISC-V computing platform with a primary focus on achieving high energy efficiency.}
\rOneHLW{The RISC-V cores of a PULP cluster can be extended, and the cluster is equipped with a heterogeneous cluster interconnect (HCI) for memory access, dataflow control, and controlling customized co-processors.}
\rOneFC{Fig.~\ref{fig:system}~(a) presents XpulpNN~\cite{garofalo2021xpulpnn}, the state-of-the-art RISC-V processor based on the PULP platform, which enables precision-scalable DNN inference with limited on-device learning capability.}
\rOneFC{By contrast, as shown in Fig.~\ref{fig:system}~(b), we propose a precision-scalable DNN processor based on the PULP platform, which significantly promotes multi-precision computing and on-device learning capability on DNNs over XpulpNN.}
\rOneFC{Compared to XpulpNN, our proposed processor stands out due to the following features:}

\rTwoHLW{\textbf{\ding{202} Resource-efficient tightly-coupled co-processor.}}
\rTwoFC{The proposed co-processor is connected with the PULP cluster through the HWPE interface, which is more resource-efficient than XpulpNN.}
\rTwoFC{The proposed co-processor achieves higher throughput by incorporating extra embedded PEs, while XpulpNN requires adding more RISC-V cores to attain improved throughput, resulting in additional associated overhead.}

\textbf{\ding{203} High-throughput systolic computing.}
\rTwoHLW{To \rtwoLQ{perform} matrix multiplication (matmul) \rtwoLQ{operations dominated in DNNs} efficiently, a high-throughput SA is proposed.}
\rTwoHLW{It requires only one instruction to initiate computation, reducing the number of compute instructions and cycles, thus increasing throughput.}
\rTwoHLW{The throughput can be further \rtwoLQ{improved} by implementing more PEs in the SA.}
\rtwoLQ{Compared to XpulpNN which requires more instructions for the same SIMD matmul and contains only four INT8 operating units per RISC-V core, the SA achieves more efficient computations and higher throughput.}

\textbf{\ding{204} Resource-efficient precision-scalable PEs.}
\rTwoFC{The constraints posed by limited hardware resources at the extreme edge make it important to improve hardware resource utilization.}
\rTwoFC{To address this, we present precision-scalable PEs that minimize resource redundancy through multiple carefully designed hardware reuse techniques.}
\rTwoHLW{Our PEs achieve \rtwoLQ{higher resource utilization} compared to XpulpNN's dotp units
\rtwoLQ{that have no reuse on multipliers supporting various precision.}}

\textbf{\ding{205} Boosted on-device learning capability.}
\rtwoLQ{To better support on-device learning at extreme-edge, floating-point computation throughput needs to be increased.}
\rtwoLQ{Therefore, our processor natively supports FP16-based computation in the PEs, rather than performing floating-point operations using the FPU embedded in the ALU, as XpulpNN does.}
\rtwoLQ{The improved floating-point computation capability enables our processor to achieve higher throughput as well as on-device learning efficiency.}

\begin{figure}[htbp]
\vspace{-11pt}
\centering
\includegraphics[width=0.45\textwidth]{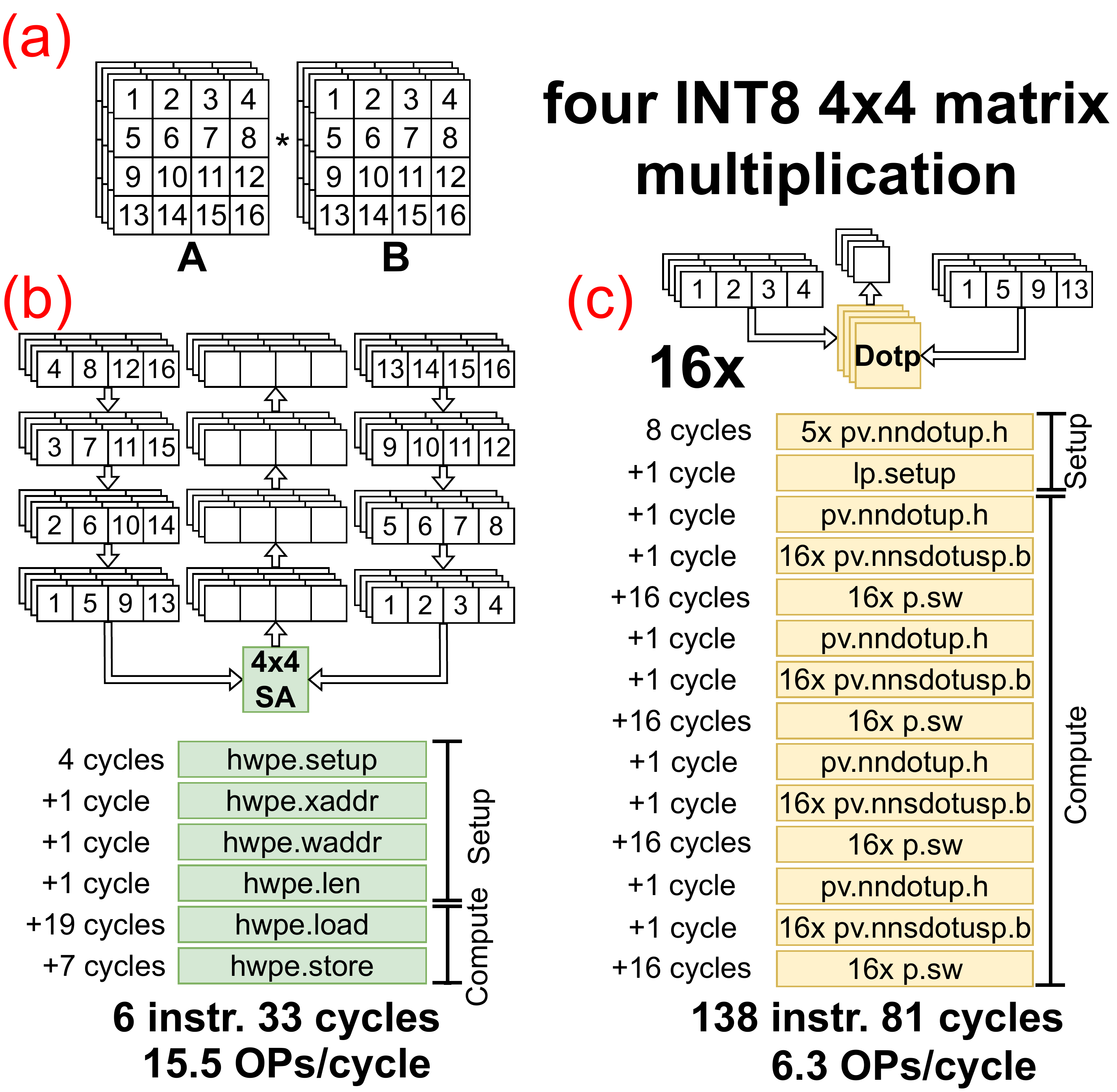}
\setlength{\abovecaptionskip}{-3pt}
\caption{\rTwoHLW{The instruction and computation flow of our processor and XpulpNN to perform an INT8 matrix multiplication operator. (a) shows the 4$\times$4 matmul operator; (b) and (c) show the computational and instruction flows of our processor's SA and XpulpNN's dotp units, respectively.}}
\label{fig:instruction_flow}
\vspace{-11pt}
\end{figure}

\subsection{Customized RISC-V Instruction-Driven Mapping}\label{subsec:mapping}
\rTwoHLW{Our processor demonstrates superior computational efficiency over XpulpNN \Review{with same multiply-accumulate (MAC) units} when executing the same matmul.
As shown in Fig.~\ref{fig:instruction_flow},
\rtwoLQ{both XpulpNN and our processor need setup and compute instructions to initialize the computation and subsequently execute operators, respectively.}
\rtwoLQ{However, our processor excels benefiting from more efficient initialization and execution.}
\rTwoHLW{For instance, it only takes 4 instructions (7 cycles) to setup, and another 2 instructions (26 cycles) to compute when employed four 4$\times$4 INT8-based operators. Our processor uses merely 4\% instructions and \Review{41\%} cycles required by XpulpNN, which needs 6 instructions (9 cycles) and 132 instructions (72 cycles) to setup and compute, respectively.}
\rtwoLQ{The significant reduction in \rTwoFC{number of} instructions and cycles contributes to a 2.5$\times$ throughput improvement for our processor, thereby achieving higher computational efficiency than XpulpNN.}}

\rtwoLQ{Specifically, a more efficient setup is first introduced to improve the computational efficiency.}
\rOneHLW{XpulpNN needs to initialize each loop when setting up the computation demonstrated in Fig.~\ref{fig:instruction_flow}~(c), whereas our processor only needs \texttt{hwpe.setup} to perform the overall initialization (including startup and precision configuration), \texttt{hwpe.xaddr} and \texttt{hwpe.waddr} to read the first addresses of input \rTwoFC{matrix} A and B\rtwoLQ{, respectively,} and \texttt{hwpe.len} to determine the data length, as shown in Fig.~\ref{fig:instruction_flow}~(b).}

\rtwoLQ{In addition, a more efficient computation is also implemented.}
\rtwoLQ{For XpulpNN, it needs multiple load and store instructions to compute all the operators, since a single dotp unit can only perform the dot-product of two vectors of length 4 per instruction, as shown in Fig.~\ref{fig:instruction_flow}~(c).}
\rTwoHLW{\rtwoLQ{By contrast,} our processor only needs one instruction to load all the data into the SA to compute (\texttt{hwpe.load}) and one instruction to store all the results (\texttt{hwpe.store}), as shown in Fig.~\ref{fig:instruction_flow}~(b).}

\begin{figure}[htbp]
\vspace{-5pt}
\centering
\includegraphics[width=0.40\textwidth]{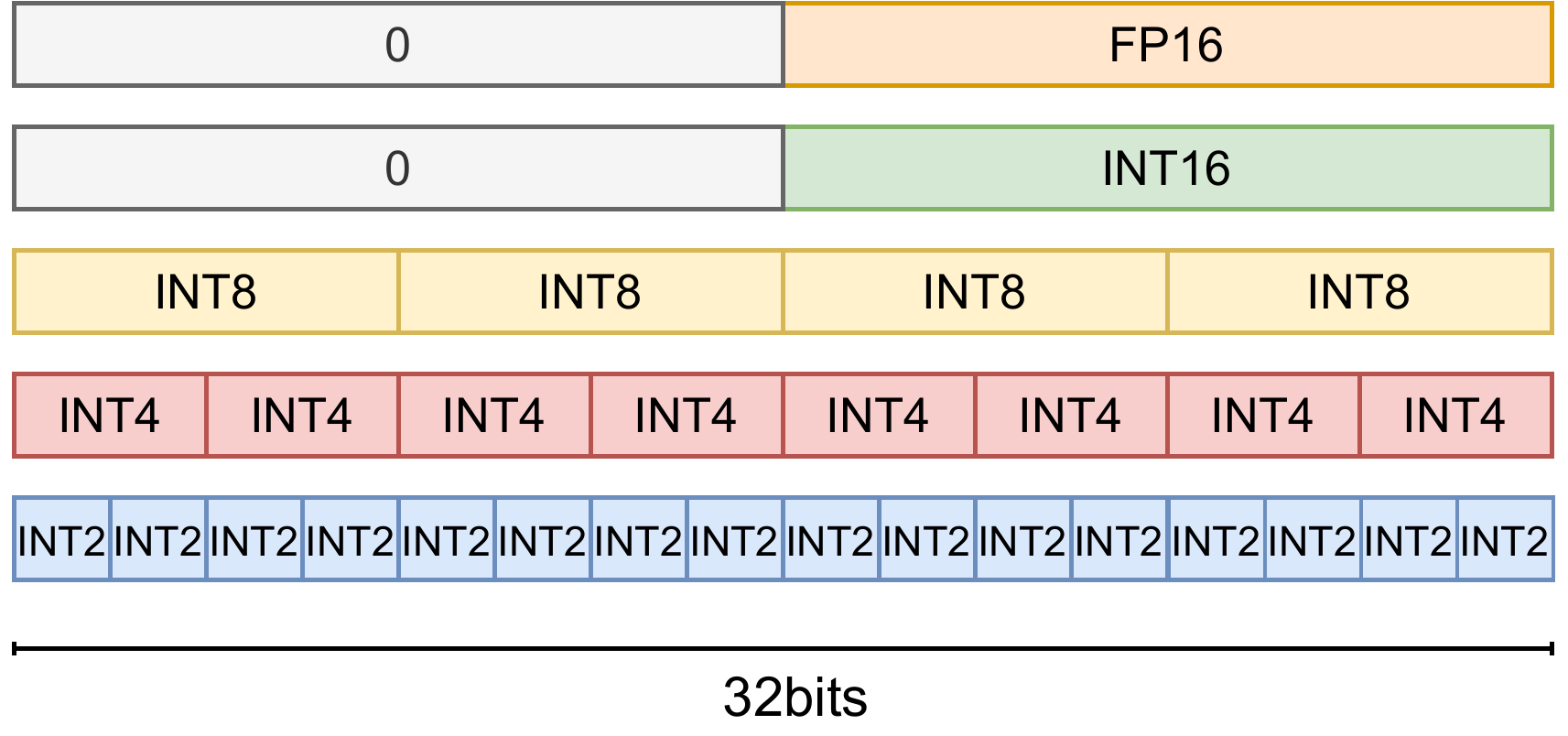}
\setlength{\abovecaptionskip}{-3pt}
\caption{Data arrangement method of different precision.}
\label{fig:data_precision}
\vspace{-10pt}
\end{figure}

\subsection{Precision-Scalable Processing Element}\label{subsec:pe}
\rTwoHLW{Fig.~\ref{fig:system} shows a precision-scalable PE utilized for highly-parallel precision-scalable MAC operations. Each PE comprises precision-scalable adder and multiplier, as well as registers and \rTwoFC{multiplexers (MUXs)} for data flow direction control. The mode of the SA determines whether the 64-bit output data Y is temporarily stored inside the PE or shifted out. During computation, 32-bit input data X and W are sequentially shifted in.}
\rTwoFC{Controlled by a precision selection signal, a PE can efficiently compute a single FP16 MAC, a single INT16 MAC, four-parallel INT8 MACs, eight-parallel INT4 MACs, or sixteen-parallel INT2 MACs on-the-fly.}

To accommodate precision-scalable PEs, a specific data arrangement method is devised to reduce overall computation time by ensuring the co-processor efficiently processes data with different precision.
\rOneLIN{Data with varying quantization length are organized in a manner depicted in Fig.~\ref{fig:data_precision}. Specifically, every four, eight, and sixteen data are grouped as a single 32-bit data in 8-bit, 4-bit, and 2-bit precision, respectively. For 16-bit INT16 and FP16 data, each of them is padded with an additional 16 bits of 0, transforming it into 32-bit data.}

\begin{figure}[htbp]
\centering
\includegraphics[width=0.45\textwidth]{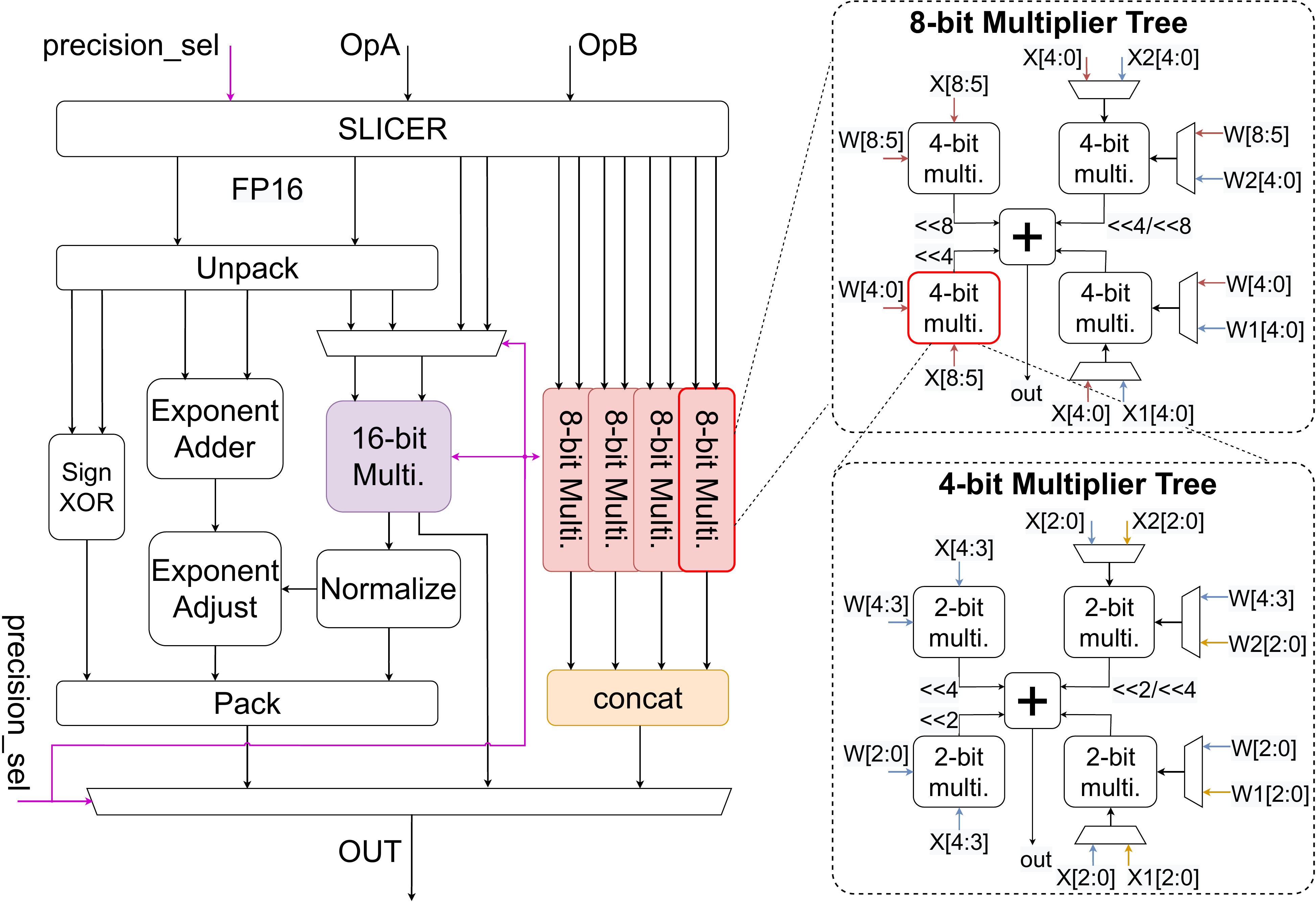}
\setlength{\abovecaptionskip}{-3pt}
\caption{\Review{Architecture of the precision-scalable multiplier with highly-reused 16-bit mantissa multiplier and 8-bit multiplier trees. Only half of the 4-bit multiplier trees and 2-bit multipliers of one 8-bit multiplier tree are reused to ensure the output bit-width remains the same at different precision levels.}}
\label{fig:multiplier}
\vspace{-4pt}
\end{figure}

\textbf{Precision-Scalable Multiplier.}
\rOneLIN{To efficiently support multi-precision INT multiplication for inference and FP16 multiplication for on-device learning,}
\rOneHLW{we propose the precision-scalable multiplier shown in Fig.~\ref{fig:multiplier}, which contains of one 16-bit multiplier and four 8-bit precision-scalable multiplier trees.}
The FP16-based mantissa multiplier reuses the hardware resources of the INT16 multiplier, while the multiplier trees are used for parallel INT8, INT4 or INT2 multiplications.
\rOneLQ{According to the precision selection signal, the inputs are sliced into different bit-widths and fed into different multipliers, and then the corresponding spliced data is selected as the outputs.}
\rOneLIN{With this precision-scalable multiplier, the hardware resources of one additional 16-bit multiplier are saved.}

\textbf{8-bit Multiplier Tree.}\label{subsubsec:tree}
The 8-bit precision-scalable multiplier tree shown in Fig.~\ref{fig:multiplier} further supports efficient computation of INT8, INT4, and INT2 multiplications. 
\Review{Each 8-bit tree consists of four 4-bit precision-scalable multiplier trees and one adder, and is capable of performing single 8-bit multiplication through shift-and-add of partial products. Different from \cite{sharma2018bit}, a single 8-bit tree also perform two 4-bit multiplications using two parallel 4-bit multiplier trees.}
Similarly, the 4-bit tree performs single INT4 multiplication or two INT2 multiplication in parallel.
This architecture saves hardware resources that would have been allocated to eight 4-bit multipliers and sixteen 2-bit multipliers, resulting in reduced LUT overhead for the PE compared to the unreused dotp unit of XpulpNN.

\begin{figure}[tbp]
\centering
\includegraphics[width=0.42\textwidth]{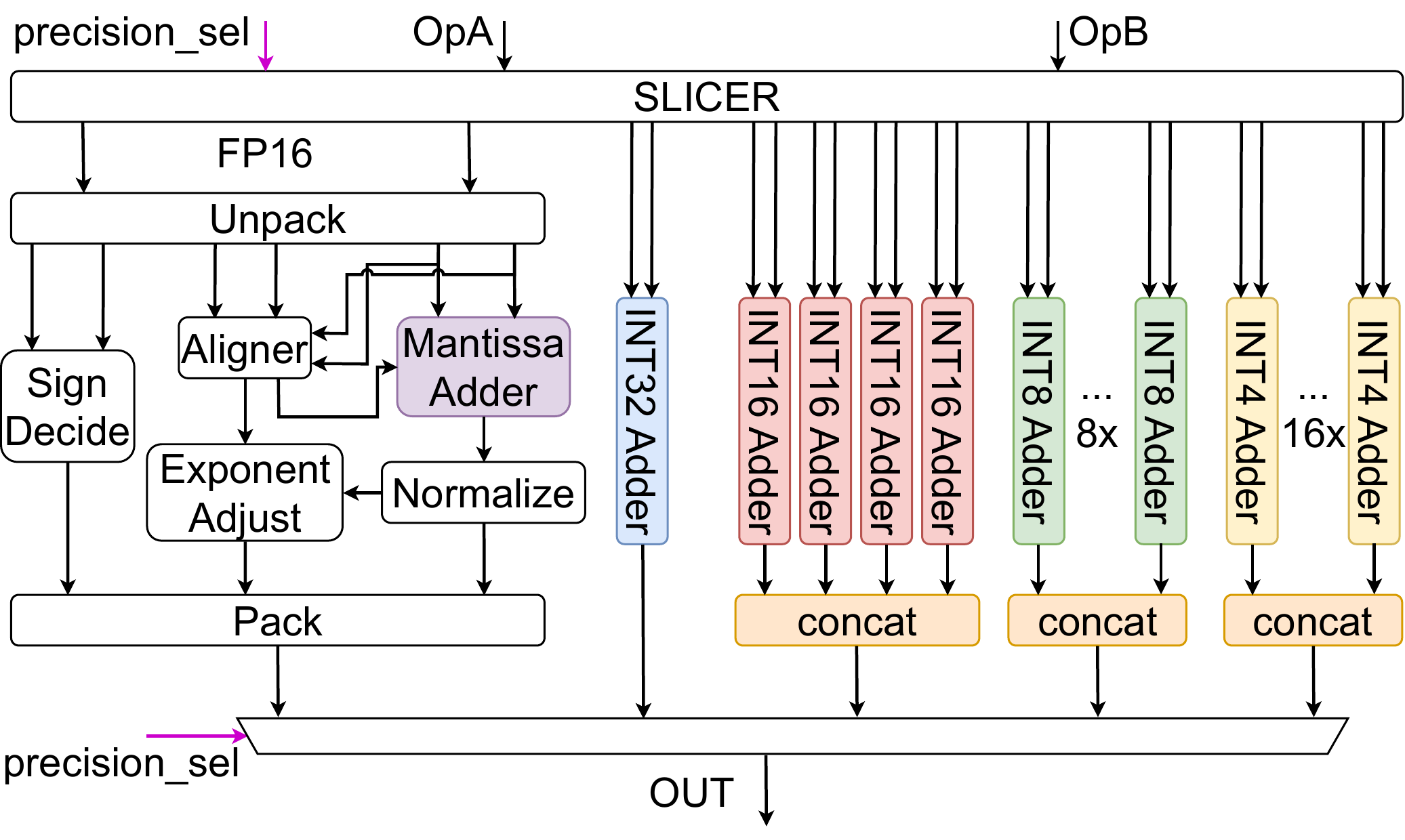}
\setlength{\abovecaptionskip}{-3pt}
\caption{Architecture of the precision-scalable adder.}
\label{fig:adder}
\vspace{-15pt}
\end{figure}

\textbf{Precision-Scalable Adder.}
In addition, we present the precision-scalable adder shown in Fig.~\ref{fig:adder} for implementing INT16, INT8, INT4, INT2 and FP16 addition to fully support precision-scalable computation and on-device learning. It consists of one FP16 floating point adder, one 32-bit adder, four 16-bit adders, eight 8-bit adders and sixteen 4-bit adders.
\rOneLQ{Unlike precision-scalable multiplier, the adder is not reused with varying precision, for that the additional overhead caused by the increased MUXs is found to be close to the hardware resources it aims to reduce.}

\subsection{Balancing LUT and DSP Mapping}\label{subsec:lut&dsp util}
\rOneLQ{For the implementation of the proposed processor architecture with FPGA devices, the computational logic is mapped to available resources including DSPs and LUTs. However, automatic resource mapping often leads to inefficiencies, such as high LUT overhead or low DSP utilization \rTwoFC{in XpulpNN}. To address this, we adopt a mapping approach to ensure high resource utilization for both DSPs and LUTs.}
In particular, the 16-bit mantissa multiplier illustrated in Fig.~\ref{fig:multiplier} is mapped to DSPs, while four 8-bit precision-scalable multiplier trees are mapped to LUTs. Moreover, the INT adders shown in Fig.~\ref{fig:adder} are mapped to DSPs, while the FP16 adder is mapped to LUTs.
\rTwoFC{This balanced mapping approach ensures efficient utilization of FPGA resources, including both DSPs and LUTs, in our design. By adopting this method, we can reduce the consumption of LUTs compared to approaches that exclusively map to LUTs, as well as relieve the over-consumption of DSPs compared to methods that primarily rely on DSP mapping.}
\section{Experimental Results}\label{sec:res}
\subsection{Experimental Setup}\label{subsec:setup}
\rOneFC{We synthesize our precision-scalable RISC-V DNN processor using Vivado 2018.3 and implement it on two FPGA boards: ZCU102 and PYNQ-Z2, with a clock frequency of 200MHz and 100MHz, respectively.}
The size of \rOneFC{SAs} in our processors deployed on ZCU102 and PYNQ-Z2 are 12$\times$12 and 4$\times$4, respectively. 
\rOneFC{The widely-used DNN models are selected for evaluation of DNN deployment, including MobileNetv2, VGG-16, ResNet-18, ResNet-50, and ViT/B-16.}
\rOneHLW{We follow the experimental methods in BARVINN\cite{askarihemmat2023barvinn} and evaluate the throughput by operators of convolutional layers and fully connected layers.}
\rOneFC{Power consumption is estimated using the Vivado Power Analysis tool.}
\rOneFC{For a fair comparison against our processor, ARM-Cortex A7 and i5-10505 are selected as CPU baselines, and Jetson Nano is selected as a GPU baseline, which is a commercial product for DNN deployment at the extreme edge.}
\rOneFC{Additionally, we implement an 8-core XpulpNN\cite{garofalo2021xpulpnn} on a ZCU102 board running at 200MHz as an FPGA baseline.}
\rOneFC{The other FPGA-based state-of-the-arts \cite{suda2016throughput, qiu2016going, chang2021mix, guo2017angel, sun2022film, askarihemmat2023barvinn} are also selected as our baselines.}

\begin{figure}[htbp]
\centering
\includegraphics[width=0.48\textwidth]{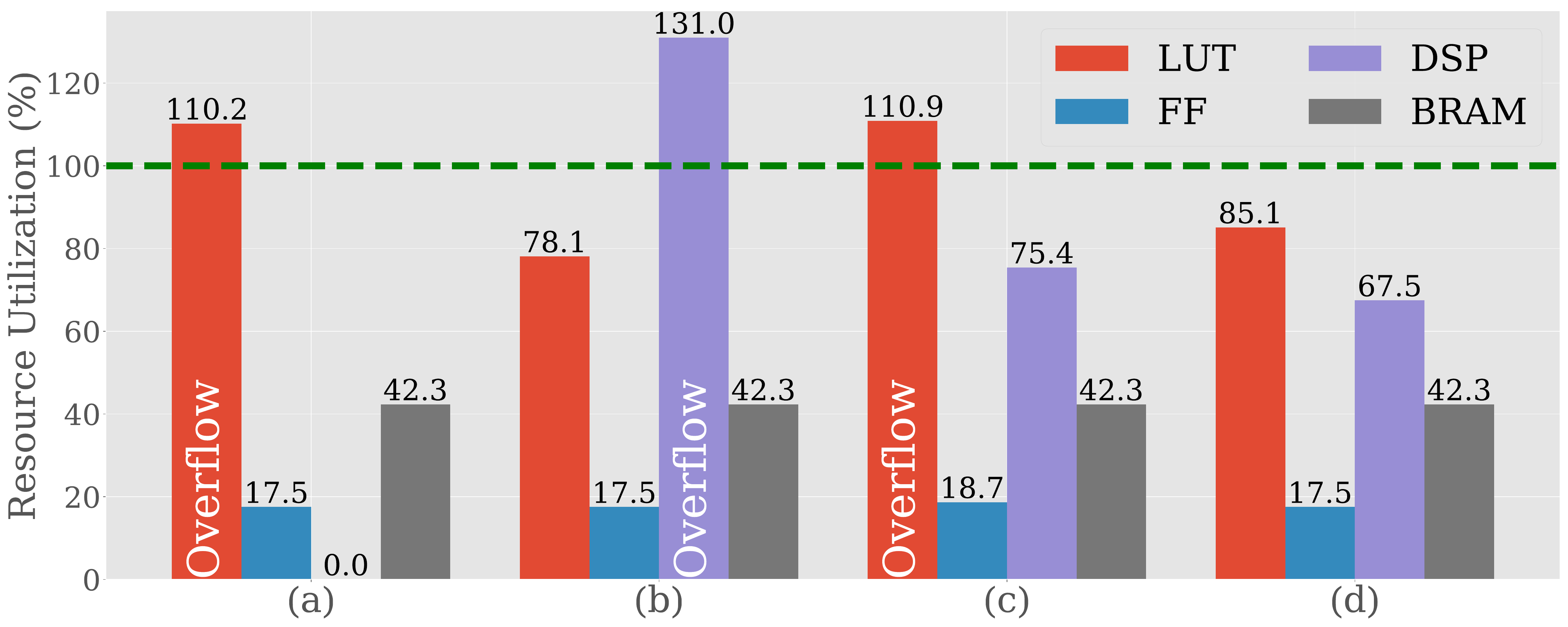}
\setlength{\abovecaptionskip}{-4pt}
\caption{Utilization using different resource mapping methods on ZCU102: (a) mapping to LUTs only with reused multipliers; (b) mapping to DSPs whenever possible with reused multipliers; (c) balanced mapping to LUTs and DSPs without reused multipliers; (d) balanced mapping to LUTs and DSPs with reused multipliers.}
\label{fig:util_r}
\vspace{-12.8pt}
\end{figure}

\subsection{FPGA Resource Utilization Analysis}\label{subsec:util}
\textbf{Reuse of Hardware Resources.}
\rOneFC{The extreme-edge platform uses FPGAs with very few on-chip resources, pressing the need to reuse hardware resources for precision-scalable inference and on-device learning.}
\rOneFC{Fig.~\ref{fig:util_r} illustrates the comparison of consumed FPGA resources on ZCU102 between the different mapping methods in \ref{subsec:pe} \& \ref{subsec:lut&dsp util}.}
\rOneHLW{Utilizing the balanced method with reused multipliers as shown in Fig.~\ref{fig:util_r}~(d) achieves 85.1\% LUT utilization and 67.5\% DSP utilization. It consumes 25.1\% fewer LUTs than mapping to LUTs only with reused multipliers in (a), 63.9\% fewer DSPs than mapping to DSPs whenever possible with reused multipliers in (b), 25.8\% fewer LUTs and 7.9\% fewer DSPs than mapping to LUTs and DSPs with the balanced method with unreused multipliers in (c). Furthermore, (d) is the only approach that does not cause resource overflow.}
\rOneLIN{These results demonstrate the effectiveness of our proposed methods in \ref{subsec:pe} \& \ref{subsec:lut&dsp util}, \rtwoLQ{which} significantly improve LUT and DSP utilization while reducing unnecessary hardware resource overhead.}

\begin{figure}[tbp]
\centering
\includegraphics[width=0.48\textwidth]{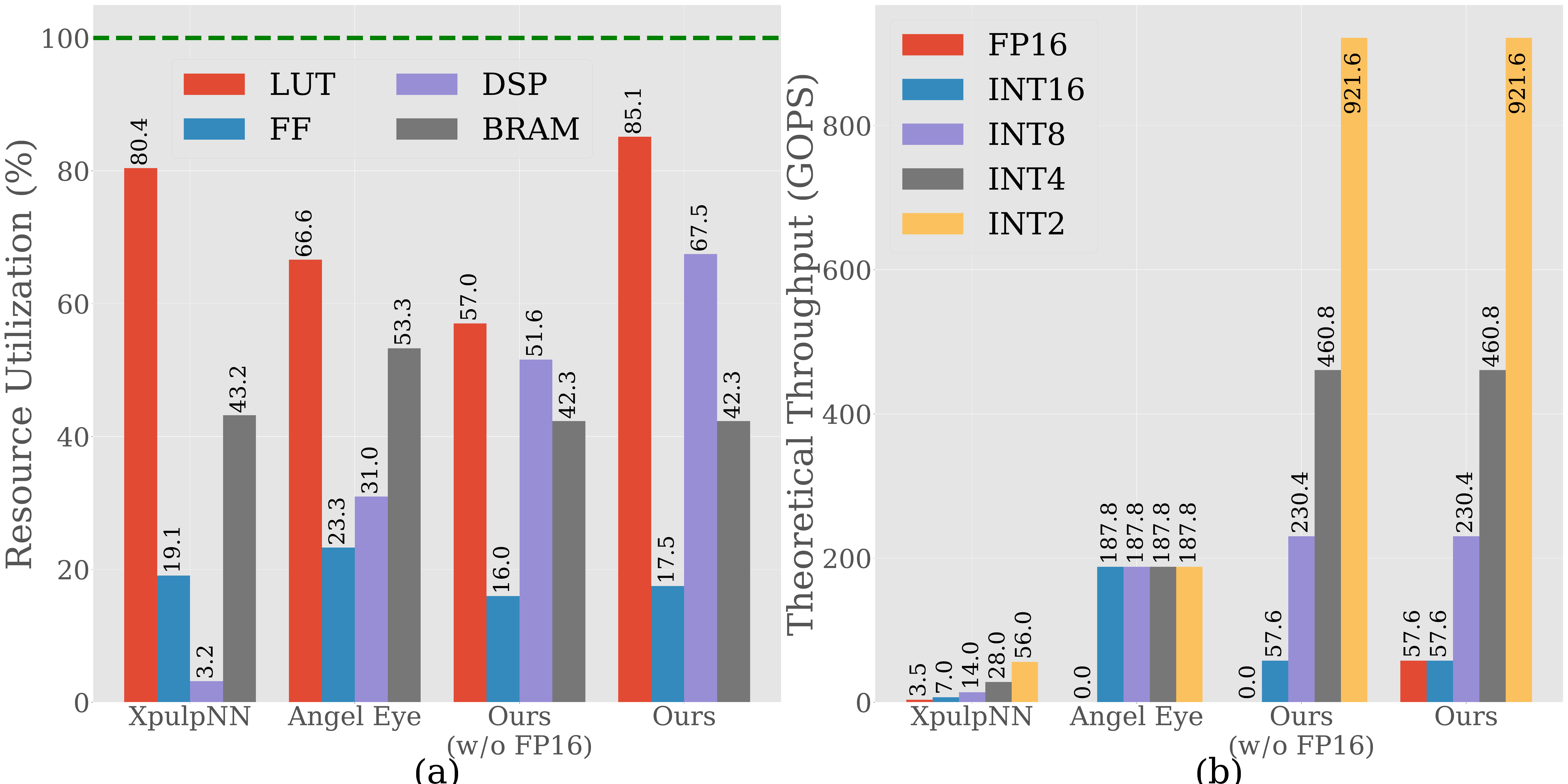}
\setlength{\abovecaptionskip}{-3pt}
\caption{Hardware resources and theoretical throughput of different works on ZCU102. Compared to XpulpNN, DSP and LUT utilization are increased by 64.3\% and 4.7\%, respectively. Furthermore, the theoretical throughput of our processor is increased by 16.5$\times$ (FP16) and 8.2$\times$ (INT8, INT4, and INT2).}
\label{fig:oltraining_resource}
\vspace{-5pt}
\end{figure}

\textbf{Overhead for On-Device Learning Support.}
\rTwoFC{The extreme-edge platform introduces additional hardware resource overhead to support floating-point computation for on-device learning.}
\rOneHLW{\rTwoFC{Hence, we evaluate} the hardware resource consumption of our processor with and without FP16 computation capability, and compare it with XpulpNN\cite{garofalo2021xpulpnn} and Angel Eye\cite{guo2017angel}, where XpulpNN supports precision-scalable together with floating-point computation, 
\rtwoLQ{and Angel Eye has decent throughput at INT16 precision despite lacking the precision-scalable capability.}}
\rOneHLW{\rtwoLQ{As shown in Fig.~\ref{fig:oltraining_resource}}, our processor has 28.1\% LUT overhead and 15.9\% DSP overhead
\rtwoLQ{when adding FP16 support,}while FF and BRAM have essentially no additional overhead. \rtwoLQ{However}, the processor achieves 57.6 GOPs of FP16 theoretical throughput.}
\rOneHLW{Compared to XpulpNN\cite{garofalo2021xpulpnn} and Angel Eye\cite{guo2017angel}, our processor increases DSP utilization \rtwoLQ{by} 64.3\% and 36.5\%, and LUT utilization \rtwoLQ{by} 4.7\% and 18.5\%, respectively. Furthermore, by supporting multi-precision and on-device learning, our processor achieves significant theoretical throughput improvement compared to XpulpNN (16.5$\times$ at FP16, 8.2$\times$ at INT8, INT4, and INT2) and Angel Eye (1.2$\times$ at INT8, 2.5$\times$ at INT4 and 4.9$\times$ at INT2).}

\begin{figure}[htbp]
\centering
\includegraphics[width=0.48\textwidth]{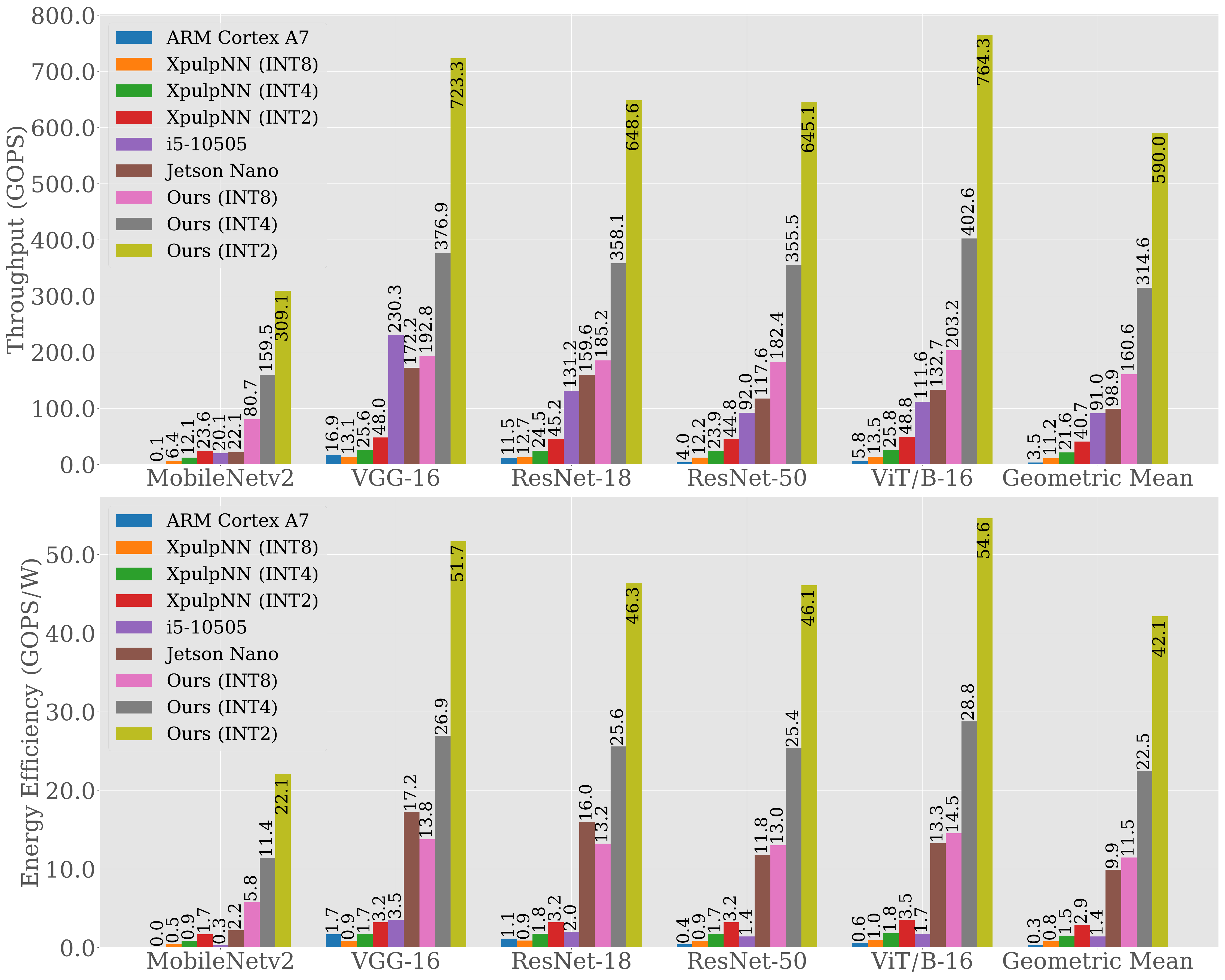}
\setlength{\abovecaptionskip}{-3pt}
\caption{\rTwoFC{Comparison of throughput and energy efficiency under various DNN models.} Our processor achieves 1.6$\times$ and 14.3$\times$ average INT8 throughput improvement and 1.2$\times$ and 14.4$\times$ average INT8 energy efficiency improvement over Jetson Nano and XpulpNN, respectively.}
\label{fig:throughput_per_model}
\vspace{-12pt}
\end{figure}

\subsection{Throughput and Energy Efficiency Comparison}\label{subsec:tp&ee}
\rTwoFC{Extreme edge platforms have stringent requirements for low DNN inference latency and are also very sensitive to power consumption. Hence, throughput and energy efficiency are critical metrics for efficient DNN inference.}
\rTwoFC{We employ our processor to deploy diverse DNN models of varying precision and make a comparison of throughput and energy efficiency between our processor and baselines.}
\rOneHLW{\rtwoLQ{The experimental results in Fig.~\ref{fig:throughput_per_model} shows} our processor achieves considerable throughput and energy efficiency on average (160.6 GOPS and 11.5 GOPS/W at INT8, 314.6 GOPS and 22.5 GOPS/W at INT4, 590.0 GOPS and 42.1 GOPS/W at INT2).}
In addition, 1.6$\times$ and 14.3$\times$ INT8 throughput improvement over Jetson Nano and XpulpNN, respectively, is achieved on average by our processor.
Furthermore, our processor demonstrates substantial advancements in energy efficiency, with an average improvement of 14.6$\times$ and 14.5$\times$ over XpulpNN for INT4 and INT2 precision, respectively.
Compared to Jetson Nano, our processor achieves 1.1$\sim$3.9$\times$ and 1.1$\sim$4.1$\times$ energy efficiency improvements on ResNet-50 and ViT/B-16 at various precision, respectively, and it achieves 2.6$\sim$10.0$\times$ higher energy efficiency improvements when running a lower computational density network, i.e., MobileNetv2, at various precision.

\begin{table*}[htbp]
\caption{Comparison to Related Works}
\label{tab:compare}
\centering
\resizebox{0.98\textwidth}{!}{%
\begin{tabular}{@{}ccccccccccccccc@{}}
\toprule
\multirow{2}{*}{Work}                  & \multirow{2}{*}{Platform} & \multirow{2}{*}{kLUTs} & \multirow{2}{*}{DSPs} & \multirow{2}{*}{\begin{tabular}[c]{@{}c@{}}Frequency\\ (MHz)\end{tabular}} & \multirow{2}{*}{Model} & \multicolumn{4}{c}{\begin{tabular}[c]{@{}c@{}}Throughput\\ (GOPS)\end{tabular}} & \multicolumn{4}{c}{\begin{tabular}[c]{@{}c@{}}Energy Efficiency\\ (GOPS/W)\end{tabular}} & \multirow{2}{*}{\begin{tabular}[c]{@{}c@{}}On-Device\\ Learning\\ FP Support\end{tabular}} \\ \cmidrule(lr){7-14}
                                       &                           &                        &                       &                                                                            &                        & INT16             & INT8               & INT4               & INT2              & INT16               & INT8                 & INT4                 & INT2                 &                                                                                       \\ \midrule
Baseline~1                            & Arm Cortex A7             & -                      & -                     & 1430                                                                       & ResNet-50              & 4.0               & N/A                  & $\times$                  & $\times$                 & 0.4                & N/A                    & $\times$                    & $\times$                    & \checkmark                                                                            \\
Baseline~2                            & i5-10505                  & -                      & -                     & 3200                                                                       & ResNet-50              & 92.0              & N/A                  & $\times$                  & $\times$                 & 1.4                & N/A                    & $\times$                    & $\times$                    & \checkmark                                                                            \\
Baseline~3                            & Jetson Nano               & -                      & -                     & 640                                                                        & ResNet-50              & N/A                 & 117.6              & $\times$                  & $\times$                 & N/A                    & 11.8                 & $\times$                    & $\times$                    & \checkmark                                                                            \\ \midrule
Going Deeper\cite{qiu2016going}        & XC7Z045                   & 218.6                  & N/A                   & 150                                                                        & VGG-16                 & 137.0               & $\times$                  & $\times$                  & $\times$                 & N/A                 & $\times$                    & $\times$                    & $\times$                    & $\times$                                                                                     \\
Angel Eye\cite{guo2017angel}           & XC7Z045                   & 182.6                  & 780                   & 150                                                                        & VGG-16                 & 187.8             & $\times$                  & $\times$                  & $\times$                 & 14.2               & $\times$                    & $\times$                    & $\times$                    & $\times$                                                                                     \\
ThroughputOpt\cite{suda2016throughput} & Stratix V                 & 153                    & 246                   & 120                                                                        & VGG-16                 & $\times$                 & 117.8              & $\times$                  & $\times$                 & $\times$                   & N/A                  & $\times$                    & $\times$                    & $\times$                                                                                     \\
Mix and Match\cite{chang2021mix}       & XC7Z045                   & 145.1                  & 900                   & 100                                                                        & ResNet-18              & $\times$                 & $\times$                  & 359.2              & $\times$                 & $\times$                   & $\times$                    & N/A                  & $\times$                    & $\times$                                                                                     \\
FILM-QNN\cite{sun2022film}             & ZCU102                    & 174.5                  & 2.1k                  & 150                                                                        & ResNet-50              & $\times$                 & N/A                & 387.8              & $\times$                 & $\times$                   & N/A                  & 28.9                 & $\times$                    & $\times$                                                                                     \\
BARVINN\cite{askarihemmat2023barvinn}  & Alveo U250                & 201.1                  & 512                   & 250                                                                        & ResNet-50              & N/A               & N/A                & N/A                & 380.4             & N/A                 & N/A                  & N/A                  & 17.7                 & $\times$                                                                                     \\
XpulpNN\cite{garofalo2021xpulpnn}      & ZCU102                    & 220.4                  & 80                    & 200                                                                        & ResNet-50              & 6.0               & 12.2               & 23.9               & 44.8              & 0.4                 & 0.9                  & 1.7                  & 3.2                  & \checkmark                                                                            \\
\textbf{Ours}                          & \textbf{PYNQ-Z2}          & \textbf{32.9}          & \textbf{190}          & \textbf{100}                                                               & \textbf{ResNet-50}     & \textbf{2.8}      & \textbf{11.8}      & \textbf{24.3}      & \textbf{46.5}     & \textbf{0.7}        & \textbf{3.0}         & \textbf{6.1}         & \textbf{11.6}        & \textbf{\checkmark}                                                                   \\
\textbf{Ours}                          & \textbf{ZCU102}           & \textbf{233.3}         & \textbf{1.7k}         & \textbf{200}                                                               & \textbf{ResNet-50}     & \textbf{47.0}     & \textbf{182.4}     & \textbf{355.5}     & \textbf{645.1}    & \textbf{3.4}        & \textbf{13.0}        & \textbf{25.4}        & \textbf{46.1}        & \textbf{\checkmark}                                                                   \\ \bottomrule
\end{tabular}%
}
\vspace{-10pt}
\end{table*}

\subsection{Comparison to CPU, GPU, and FPGA-based Prior Arts}\label{subsec:comp}
\rOneHLW{To evaluate the improvement of our processor on various metrics, we make a comprehensive comparison with CPUs, GPU, and FPGA-based prior works, as detailed in Table~\ref{tab:compare}.}
\rTwoFC{Compared to CPUs such as i5-10505 and Arm Cortex A7, our processor achieves noteworthy energy efficiency enhancements of 2.4$\times$ and 8.5$\times$, respectively, at INT16 precision on ZCU102. Additionally, it outperforms the energy-efficient GPU, Jetson Nano, with a throughput improvement of 1.6$\times$ and an energy efficiency improvement of 1.1$\times$.}
\rOneHLW{When compared to FPGA-based XpulpNN\cite{garofalo2021xpulpnn} that supports precision-scalable inference and on-device learning, significant improvements in throughput by 7.8$\sim$15.0$\times$ and energy efficiency by 8.5$\sim$14.4$\times$ have been achieved by our processor on ZCU102 at different precision.}
\rTwoFC{Our processor uniquely offers both precision scalability ranging from INT16 to INT2 and on-device learning capability with FP16 support, setting it apart from prior FPGA-based arts\cite{suda2016throughput, chang2021mix, guo2017angel, qiu2016going, askarihemmat2023barvinn, sun2022film}. Furthermore, our processor on ZCU102 improves throughput by 1.6$\sim$1.7$\times$ when compared to these works.}
\rOneLQ{Finally, our processor can accommodate various FPGA platforms with different amount of hardware resources. When deployed on PYNQ-Z2, it achieves comparable throughput and 1.8$\sim$3.6$\times$ energy efficiency gains compared to XpulpNN at INT8, INT4, and INT2 precision.}
\vspace{-1.45pt}
\section{Conclusion}\label{sec:concls}
In this paper, we propose a high-throughput, energy-efficient and precision-scalable RISC-V DNN processor with on-device learning capability at the extreme edge to address the growing needs for DNN deployment.
Furthermore, precision-scalable processing elements with boosted on-device learning throughput are proposed, and the hardware resources are fully leveraged by several methods.
Our proposed processor increases LUT and DSP utilization by 4.7\% and 64.3\%, respectively, and achieves up to 14.6$\times$ improvement of inference throughput and energy efficiency on average compared to the existing XpulpNN solutions when computing various quantized DNNs. It also achieves up to 16.5$\times$ on-device learning throughput improvement compared to XpulpNN.

%
\IEEEpeerreviewmaketitle


\section*{Acknowledgment}

This work was supported in part by the National Key R\&D Program of China under Grant 2022YFB4400604, in part by the National Natural Science Foundation of China under Grant 62174084, and in part by the Postgraduate Research \& Practice Innovation Program of Jiangsu Province under Grant SJCX23\_0016.



%

\bibliographystyle{IEEEtran}
\bibliography{ref}	

\end{document}